# Shedding light on the monolayer-WSe$_2$ exciton's nature by optical effective-mass measurements

Lorenz Maximilian Schneider[1], Shanece S. Esdaille[2], Daniel A. Rhodes[2], Katayun Barmak[3], James C. Hone[2,*], and Arash Rahimi-Iman[1,*]

[1]*Faculty of Physics and Materials Sciences Center, Philipps-Universität Marburg, Marburg, 35032, Germany*

[2]*Department of Mechanical Engineering, Columbia University, New York, NY 10027, USA*

[3]*Department of Applied Physics and Applied Math, Columbia University, New York, NY 10027, USA*

**<Abstract>**

Two-dimensional excitons formed in quantum materials such as monolayer transition-metal dichalcogenides and their strong light–matter interaction have attracted unrivalled attention by the research community due to their extraordinarily large oscillator strength as well as binding energy, and the inherent spin–valley locking[1,2]. Semiconducting few-layer and monolayer materials with their sharp optical resonances such as WSe$_2$ have been extensively studied and envisioned for applications in the weak[3–5] as well as strong light–matter coupling[6–9] regimes, for effective nano-laser operation with various different structures[10–14], and particularly for valleytronic nanophotonics motivated by the circular dichroism[15,16]. Many of these applications, which may benefit heavily from the two-dimensional electronic quasiparticle's properties in such films, require controlling, manipulating and first of all understanding the nature of the optical resonances that are attributed to exciton modes. While theory[17–19] and previous experiments[20–23] have provided unique methods to the characterization and classification efforts regarding the band structure and optical modes in 2D materials, here, we directly measure the quasiparticle energy–momentum dispersion for the first time. Our results for single-layer WSe$_2$ clearly indicate an emission regime predominantly governed by free excitons, i.e. Coulomb-bound electron–hole pairs with centre-of-mass momentum and corresponding effective mass. Besides uniquely evidencing the existence of free excitons at cryogenic temperatures optically, the fading of the dispersive character for increased temperatures or excitation densities reveals a transition to a regime with profound role of charge-carrier plasma or localized excitons regarding its emission, debunking the myth of free-exciton emission at elevated temperatures.

**</Abstract>**

**\<Text body\>**

In solids, fundamental excitation of matter is characterized by the formation of hydrogen-like quasiparticles[24] consisting of an excited electron Coulomb-bound to a defect electron—a "hole" in the valence band. The optical transition of this quasiparticle state, which is referred to as exciton, is less energetic than the direct band-gap transition, due to the binding energy of the electronically-neutral bound system, which can be described in the two-particle picture with a free-particle dispersion and effective mass (see **Fig. 1a**). The resulting "optical gap" is very characteristic for absorption and emission properties of semiconductor quantum structures[25–27] particularly in the case of two-dimensional (2D) transition-metal dichalcogenides (TMDCs) with extraordinary binding energies on the order of 0.5 eV in the monolayer (ML) regime[19,20,28,29] owing to the strong quantum confinement and optical band gaps in the visible to near-infrared spectral region (see Ref. [[30]] and references therein). Monolayer TMDCs such as WSe$_2$, furthermore, feature a direct band gap configuration at two equivalent time-symmetry-inverted momentum-space valleys, K and K', and a strong oscillator strength of electron–hole pairs. On top, spin–valley locking gives rise to circular dichroism and helicity-dependent light–matter interaction with the respective valleys. Comprehensibly, owing to the fact that room temperature resonances resemble excitonic behaviour with considerable binding energies, these materials have become very attractive for various applications ranging from photonics[30] to valleytronics[15,16,31].

Owing to the fact that emission out of the exciton line does not necessarily correspond to the decay of the expected Coulomb-bound particles[32], care needs to be taken when interpreting signatures that can be attributed to exciton modes. It is understood that the presence of exciton resonances lets the electron–hole plasma excited in semiconductors radiate out of the exciton mode. While the build-up of an (incoherent) 2D-exciton population on ps-to-ns timescales can be probed effectively using ultrafast optical-pump–THz/IR-probe schemes, which can cover the intra-excitonic transitions (see Ref. [[33]]), as a means of verification of formed quasiparticles and timescale investigations[22], density-dependent probing of the absorption edge is used for the investigation of a transition from a Mott-insulator state to an electron–hole-plasma with eyes towards population inversion and band-gap renormalization[21]. However, since hydrogen-like transitions and density-dependent binding-energy reduction or bleaching in the material are not unambiguous signatures for the presence of free excitons in monolayers, but clear features of excitonic populations, their presence—next to the occurrence of bound/localized excitons and electron–hole plasma—can so far only be indirectly evidenced by charge-carrier-density-dependent photoluminescence studies where a linearity factor in the intensity increase = 1 indicates free excitons[34–36]—in contrast to <1 for bound electrons and holes. Such investigations can indeed give a hint at the transition from a free to a bound exciton regime and from excitonic to plasma emission. Previously, temperature-dependent studies on monolayer WSe$_2$ involving also hexagonal-boron-nitride- (h-BN) encapsulated monolayers indicated such transition to a bound-excitonic regime by a temperature-dependent decrease of the linearity factor above 100 K[37]. Similarly, a kink in the input-output linearity factor at higher excitation densities near to the Mott transition is found for TMDCs caused by a different band renormalization at K and Σ-point[38,39].

Here, we demonstrate the first-of-its-kind direct optical measurement of the free-exciton dispersion—with parabolic centre-of-mass-momentum dependence—in a 2D semiconductor, whereby an effective exciton mass $m^*$ at cryogenic temperatures and low excitation densities can be extracted for WSe$_2$ that is in good agreement with the published theoretical[18,40] and experimentally-deduced values[41]. It is worth noting that such dispersion measurements have not even been performed explicitly for ordinary planar semiconductor quantum structures owing to the expected shallow curvatures in the case of typical effective quantum-well-exciton masses ~>0.5 $m_e$ (electron mass). Furthermore, we investigate the expected behaviour from a 2D-confined freely-propagating Coulomb-bound (Wannier-like[42]) two-particle system with in-plane momentum dispersion as a function of temperature, excitation densities, and in relation to quasi-resonant and off-resonant excitation.

In our cryogenic optical spectroscopy measurements, a temperature-dependent transition from a free-exciton to a bound-exciton situation above 100 K is indicated. We also see an excitation-density-dependent increase of the effective mass at elevated densities towards an intermediate regime, in which a free exciton gas and an electron–hole plasma are present at the same time, by an effective increase of unbound electron–hole pairs. Ultimately, choosing an off-resonant above-band-gap injection of hot carriers as excitation scheme, the measured dispersion becomes flat and represents a plasma-dominated emission at the exciton resonance. These changes in the energy–momentum dispersion are in line with expectations for the three regimes and highlight the potential of our experiment, which set stringent requirements to the optical setup and the sample in order to be successful.

For a free 2D-exciton with in-plane centre-of-mass motion, its energy can be described by

$$(1) \qquad E^{2D}(\boldsymbol{K}) = E_0(\boldsymbol{k_0} = 0) + \frac{\hbar^2 \boldsymbol{q}_{||}^2}{2m^*}$$

with $E_0(\boldsymbol{k_0})$ being the ground-state energy at zero in-plane momentum $\boldsymbol{q}_{||}$ and a kinetic-energy term quadratic in $\boldsymbol{q}_{||}$ known for free particles with effective mass $m^*$. Owing to this dispersion relation and energy as well as momentum conservation, emitted photons resulting from the radiative collapse of free excitons of certain mass and momentum will carry information about the particle via its own momentum and energy (schematically shown in **Fig. 1a**). Decomposing the wave vector **k** of the photon into an out-of-plane $k_0$ and in-plane component $k_{||}$, the angle of emission in vacuum $\theta_0$ with respect to the normal is directly linked to the transverse momentum of the photon $k_{||}$ via the following equation:

$$(2) \qquad k_{||} = \frac{2\pi n}{\lambda_0} \tan\left(arcsin\left(\frac{sin(\theta_0)}{n}\right)\right),$$

with *n* the refractive index of the medium and $|k_0|=2\pi/\lambda_0$ the vacuum wave-vector of ground-state emission. In the case of emission perpendicular to the 2D film, photons exhibit the energy $E_0(k_0) = \frac{hc|k_0|}{2\pi} = \frac{hc}{\lambda_0} = \hbar\omega_0$ of the free quasiparticle's ground-state, with *h* the Planck constant, *c* the speed of light and $\lambda_0$ the wavelength in vacuum. Thus, angle-resolved spectroscopy can be utilized to access such energy–momentum dispersion relation.

To achieve the desired signatures experimentally, care has to be taken with the choice of the measurement parameters. For a maximized optical collection of emitted angles, which correspond to the in-plane momentum by the **Equation (2)**, a numerical aperture NA = 0.6 is chosen for the employed microscope objective, while the working distance must enable sample studies behind a cryostat window. By Fourier-space projection, the imaging optics of the setup project angles up to $\theta_0 = \pm37°$ onto the imaging monochromator's entrance-slit plane (see **Methods** for further details). In contrast to angle-resolved photo-electron spectroscopy (ARPES), which resolves the electronic band structure, angle-resolved photoluminescence spectroscopy provides an incredibly direct, simple and reliable access to the excitonic dispersion. However, due to typically heavy effective masses of excitons and the fact that the optically accessible part of the dispersion is very close to the zero-momentum point, the parabolic feature with detectable angle-dependent energy changes in Fourier-space-resolved spectra on the order of (sub-)meV can only be observed for very narrow emission lines and sufficiently high optical resolution in *E* and **k**.

For our investigations, an h-BN sandwiched high-quality WSe$_2$ monolayer on an SiO$_2$/Si substrate (see **Fig. 1b**) was mounted in an optical micro-cryostat for spectroscopy at temperatures down to 10 K. For details on the stacking procedure we refer to the **Methods** section and the **Supporting Information** for micrographs. From such high-quality sandwich stack, a very narrow linewidth of the excitonic photoluminescence is obtained at cryogenic temperatures (**Fig. 1c**) which is essential for the resolution of the energy-momentum dispersion in practical measurements (**Fig. 1d**).

An excitation-density-dependent study clearly shows a curved dispersion of the WSe$_2$ exciton mode which is reduced in curvature at elevated densities. **Figure 2a** and **b** depict dispersion curves measured at 15 K for the excitation densities 0.1 and 3.6 kW/cm², respectively. Here, even at the highest (safe-to-pump) power density, a parabolic behaviour can be deduced using the simple free-particle in-plane dispersion relation with effective masses (**Equation (1)**). Dashed lines represent fitted parabolic dispersion curves corresponding to the extracted effective masses. **Figure 2c** shows an overview on the density-dependent changes obtained in this density range, showing a reduction of the curvature that corresponds to an increase of the effective mass $m^* = m_{eff} \, m_e$ from 0.25 to 0.28 $m_e$ as labelled in the plot. This trend can be expected for a transition from a regime with (negligibly) low fraction of electron–hole plasma to one with an increased fraction, leading to higher optically-retrieved effective masses at higher pump-densities under quasi-resonant excitation conditions. Here, the obtained effective mass at 15 K and low pump density well matches the theoretically expected value of 0.29 $m_e$ for free excitons in encapsulated monolayer WSe$_2$[18].

Remarkably, but not surprisingly, the excitation conditions strongly influence the emission regime obtained for the 2D system. While near-resonant optical excitation with the Titanium-Sapphire laser in continuous-wave mode enabled us to detect dispersion characteristics typical and realistic for free excitons in our 2D-material system, high-energy off-resonant excitation with a blue continuous-wave laser diode led to a completely lost angle-dependence of the emission out of the exciton line.

The choice of these two excitation modes had to serve experimental necessities. On the one hand, strong suppression of the laser light in optical spectra at close-to-resonant excitation conditions below the electronic band gap were needed which was achieved only for narrow-line continuous-wave operation together with a high-performance long-pass filter. On the other hand, blue light was needed for strictly off-resonant laser wavelengths beyond the tunability range of the Titanium-Sapphire laser. We refer to the **Methods** for details. A comparison of the Fourier-space spectra at about 15 K for both excitation schemes is shown in the **Supporting Information**.

It is important to note that continuous-wave pumping provides steady-state photoluminescence from the investigated excitonic mode at equilibrium conditions of the electronic and lattice system. Thereby, any time-resolved density-dependent effects, which could arise from population build-up, relaxation and cooling processes and temporal heat dissipation into the lattice, are ruled out. Moreover, intensity levels for time-integrated acquisition are good enough under continuous-wave pumping for imaging spectroscopy probing the Fourier space. In contrast, time-integrated spectroscopy with pulsed irradiation would lead to a smearing out of the exciton dispersion due to any dynamics in the system, as every single time window in the ps scale would constitute a different excitation regime.

Free-exciton emission characteristics are unambiguously obtained in temperature-dependent optical dispersion measurements for temperatures up to 100 K. **Figure 3a** shows an energy–momentum dispersion measured for monolayer WSe$_2$ at 15 K for quasi-resonant pumping at 3.6 kW/cm². Similar data for 100 and 150 K is shown in **Figure 3b** and **c**, respectively. An emission spectrum measured at 100 K with off-resonant pumping at 2 kW/cm$^2$ is shown in **Figure 3d**, which features a flat dispersion in contrast to the quasi-resonant excitation scenario in **b**.

While non-resonantly excited emission exhibits no curvature at all, the unambiguous occurrence of a curved dispersion up to about 100 K for quasi-resonant excitation is attributed to the presence of free excitons below 100 K. However, above 100 K, the recorded photoluminescence signal cannot be attributed to radiative recombination of free excitons even for quasi-resonant excitation, i.e. the presence of free excitons cannot be asserted.

The disappearance of the free-exciton signature ultimately sheds light on the nature of those species, which are observed at elevated temperatures. This finding improves our general understanding to that extent that excitonic species at room temperature in monolayer WSe$_2$ (and likely all other 2D semiconductors of the same family) cannot be thought of as free excitons, but have to be regarded as bound/localized excitons or plasma. The loss of curvature for angle-resolved data for off-resonant pumping, however, can be explained by the electron–hole-plasma dominance in the monolayer's emission.

**Figure 3e** summarizes the temperature series for quasi-resonant excitation. The error bars of the dispersion data points (also in **Figure 2c**) correspond to the Gauss-fit uncertainty of the wavelength-resolved line fits. Dashed lines represent fitted parabolic dispersion curves corresponding to the extracted effective masses labelling the data.

In contrast to the free exciton with centre-of-mass propagation in the plane of 2D materials, which are supposed to exhibit a measurable dispersion in agreement with theoretically predicted effective masses, electron-hole plasma emission out of the exciton resonance cannot exhibit any dispersion according to theoretical considerations. However, optical recombination for unbound electron-hole pairs takes statistically place for all possible angles at the detected energies in consideration of the equations below after Ref. [32]. **Equation (3)** is a steady-state solution to the semiconductor luminescence equations (SLE), a commonly used many-body approach for modelling emission of semiconductors that is valid for a low excitation density. It describes the PL intensity at energy $E = \hbar\omega$ and momentum **q**:

$$(3) \qquad I_{PL}(\omega_q) = \frac{2F_q^2}{\hbar} Im \left[ \sum_\lambda \frac{|\phi_\lambda^R(r=0)|^2 \left( \Delta N_\lambda^{exc}(q) + N_\lambda^{eh}(q) \right)}{E_\lambda - \hbar\omega_q - i\gamma} \right]$$

$$(4) \qquad N_\lambda^{eh}(\boldsymbol{q}) \equiv \sum_{\boldsymbol{k}} \phi_\lambda^L(\boldsymbol{k})^2 f_{\boldsymbol{k}+q_e}^e f_{\boldsymbol{k}-q_h}^h$$

with $\boldsymbol{q_e} + \boldsymbol{q_h} = \boldsymbol{q}$ and $F$ the matrix element of light matter coupling, $\phi_\lambda^R$ the right-handed eigen-function of the excitonic state $\lambda$ (1s, 2s, etc.) with energy $E_\lambda$ in real space coordinates. $\phi_\lambda^L$ corresponds to the left-handed eigen-function, respectively. $\Delta N_\lambda^{exc}(\boldsymbol{q})$ and $N_\lambda^{eh}(\boldsymbol{q})$ are the sources of PL, namely the excitonic part and emission from free carriers. $f_k^e$ and $f_k^h$ are the populations of electrons and holes, respectively, with momentum $\boldsymbol{k}$, as described by the semiconductor Bloch equations (SBE) and $\gamma$ is a phenomenological dephasing term. **Equation (4)** shows that the emission of plasma with momentum $\boldsymbol{q}$ is not linked to the momentum of the carriers $\boldsymbol{k}$, but to the difference of momentum of the hole and electron. Thus, no dispersion can be measured statistically.

These theoretical considerations explain the lack of dispersion in angle-resolved measurements for off-resonant pumping schemes. Continuous-wave quasi-resonant pumping, however, directly leads to the matter excitation corresponding to exciton formation and is evidenced by the optically-measured parabolic dispersions in time-integrated spectra representing the system in its steady-state configuration.

In conclusion, the presented work has focused on energy–momentum dispersion studies for monolayer WSe$_2$ using our direct optical measurements of Fourier-space resolved spectra, which confirm the presence of free excitons with their effective mass at cryogenic temperatures. Moreover, we also show a density- and temperature-dependent transition from a free-quasiparticle-dominated regime to an intermediate phase with plasma and to bound-state emission, respectively, for an archetype 2D material. Changes of the energy–momentum dispersions' curvatures were unambiguously and astonishingly measured via angle-resolved photoluminescence for high-quality WSe$_2$ encapsulated between thin h-BN providing excellent excitonic linewidth conditions for such an unrivalled study.

**</Text body>**

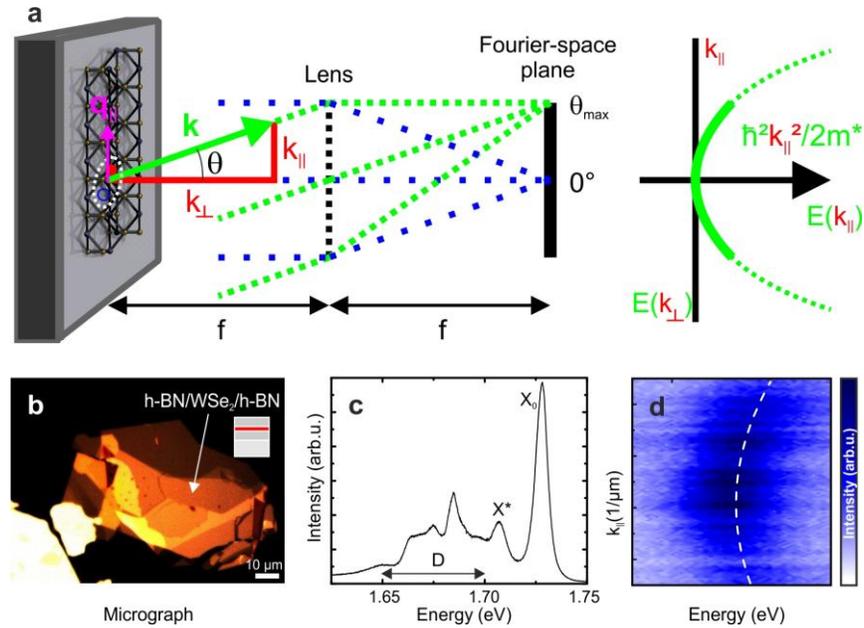

**Figure 1 | Schematics, micrograph of the sample and optical properties. a** *Schematic exciton propagation in a 2D lattice with in-plane momentum* **q**$_{||}$ *and emission of photons with wave vector* **k**, *with the corresponding projection of detected angles in the Fourier space and corresponding in-plane dispersion relation.* **b** *Microscopy image of the h-BN-WSe$_2$-h-BN stack on an SiO$_2$/Si substrate. The sandwiched monolayer region is indicated by an arrow. Images taken during the stacking sequence and detailed markings are found in the Supporting Information. Inset: Schematic 2D heterostructure.* **c** *Optical spectrum of the encapsulated high-quality monolayer with exciton X$_0$, trion X* *and defect states D.* **d** *Exemplary k-space-resolved spectrum of a monolayer free-exciton mode at 15 K.*

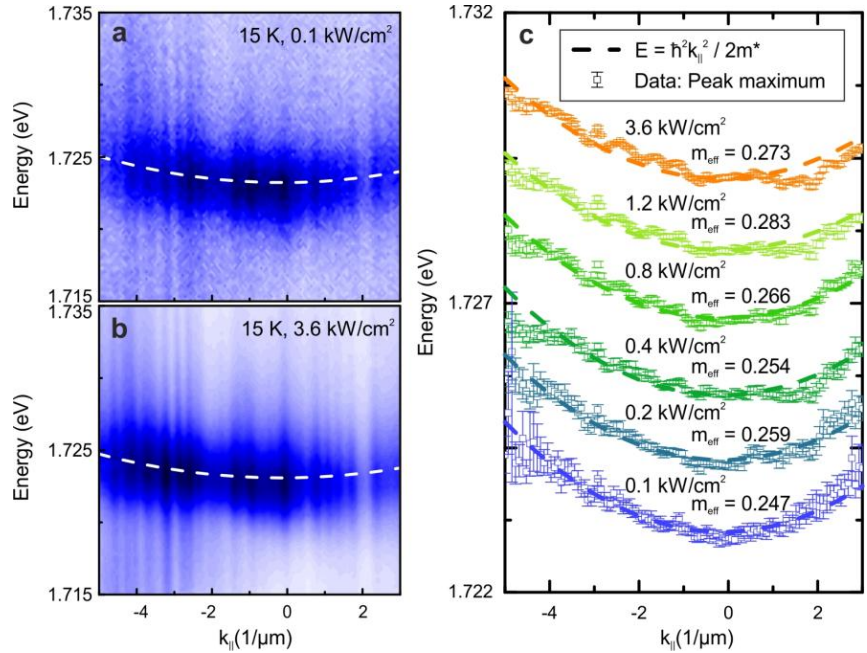

**Figure 2 | Density-dependent dispersion changes of the WSe₂ exciton mode. a** and **b** depict energy–momentum-dispersion curves measured at 15 K in false-colour intensity representation in arbitrary units (linear colour scale from white to dark blue corresponding to minimum to maximum signal), for the excitation densities 0.1 and 3.6 kW/cm², respectively. **c** An overview on the density-dependent changes is given for the measurement range, showing a reduction of the curvature that corresponds to higher effective masses. Dashed lines represent fitted parabolic dispersion curves corresponding to the extracted effective masses labelling the data. Dispersion data is offset vertically by 1.1 meV for clarity. Error bars represent fit errors of Gaussian fits to the individual line spectra.

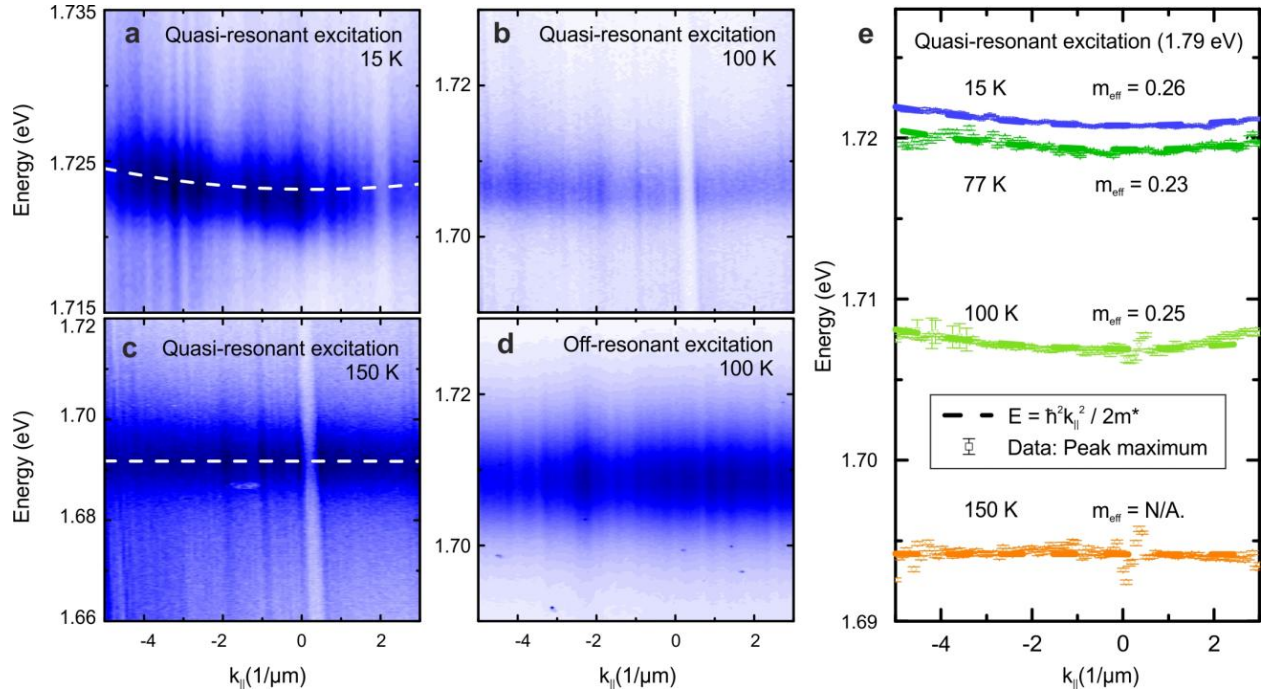

**Figure 3 | Free-exciton emission characteristics in temperature-dependent optical dispersion measurements. a** *Energy–momentum dispersion (false-colour intensity representation, similar to Fig. 2) for monolayer WSe₂ obtained at 15 K for quasi-resonant pumping. Similar data for 100 and 150 K is shown in **b** and **c**, respectively. For comparison to the corresponding quasi-resonant case in **b**, an emission spectrum measured at 100 K with off-resonant pumping is displayed in **d**. **e** summarizes the temperature series for quasi-resonant excitation at 15 K, 77 K, 100 K and 150 K. Error bars represent fit errors of Gaussian fits to the individual line spectra. Dashed lines represent fitted parabolic dispersion curves corresponding to the extracted effective masses labelling the data.*

**Methods**

**Sample preparation.** Tungsten diselenide (WSe$_2$) bulk single crystals were grown in an excess selenium flux (defect density: 5 x 10$^{10}$/cm$^2$) (see Ref. [43]). For encapsulated samples, monolayer WSe$_2$ and h-BN were first exfoliated from bulk single crystals onto SiO$_2$. For WSe$_2$, the SiO$_2$ substrate was first exposed to an O$_2$ plasma step before exfoliation. Monolayers and thin h-BN were both identified by optical contrast using a microscope. Afterwards, a dry stacking technique using polypropylene carbonate (PPC) on PDMS was used to pickup and stack h-BN/WSe$_2$ layers. First a top layer of h-BN (~20 nm) is picked up at 48 degrees C, then WSe$_2$, and finally the bottom layer of h-BN (~30 nm). After each h-BN pickup step the PPC is briefly heated to 90 C to re-smooth the PPC and ensure a clean wave front. For transferring the stack onto a clean substrate, the substrate is first heated to 75 degrees C, the stack is then put into contact, and gradually heated to 120 degrees C. Afterwards, the PPC/PDMS is lifted and the substrate is immersed in chloroform and rinsed with IPA to remove polymer residue.

**Optical measurement.** Angle-resolved micro-photoluminescence (μ-PL) measurements were performed using a self-built confocal optical microscope with Fourier-space imaging capabilities (similar to Refs. [44,45]). The sample was mounted in an evacuated (~10$^{-7}$ mbar) helium-flow cryostat, which was placed under the 40x microscope objective of the μ-PL setup. All data shown are time-integrated spectra. For quasi-resonant excitation, a continuous-wave-operated titanium-sapphire laser (*SpectraPhysics Tsunami)* was used emitting light at 693 nm. A high aspect-ratio long-pass filter with specified edge at 700 nm was used to block the laser light. Off-resonant pumping was achieved with a 445-nm continuous-wave diode laser. For detection, a nitrogen-cooled charge-coupled device (CCD) behind an imaging monochromator (*Princeton Instruments Acton SP2300*) was used, using two-dimensional chip read-out. Dispersion curves that are intensity (counts per integration time) over wavelength (long axis) over momentum (short axis) were recorded via the exposed CCD area. For *k*-space-resolved spectroscopy, the Fourier-space image of the sample's emission is projected onto the monochromator entrance slit via a set of lenses. Effective-mass fit certainty is 0.01 - 0.02 $m_e$. The investigated excitation densities were set by a neutral density filter wheel (discrete steps). Detection of the μ-PL signal from the sample took place behind a spatially-filtering aperture in the real-space projection plane of the confocal microscope, which selects a spot of ~1 μm diameter. The laser-spot diameter amounts to approximately 2 μm.

**Acknowledgement**

The authors acknowledge financial support by the German Research Foundation (DFG: SFB1083, RA2841/5-1), by the Philipps-Universität Marburg, and the German Academic Exchange Service (DAAD). Synthesis of WSe$_2$ and heterostructure assembly are supported by the NSF MRSEC program through Columbia in the Center for Precision Assembly of Superstratic and Superatomic Solids (DMR-1420634).

**Authors' contributions**

A.R.-I. conceived the experiment and initiated the study on 2D-exciton-dispersion measurements in 2015. A.R.-I. guided the joint work together with J.C.H. High-quality WSe$_2$ synthesis and heterostructure assembly were achieved by S.S.E., D.A.R., K.B. and J.C.H. The setup was established by L.M.S. and A.R.-I., and the structures measured by L.M.S. The results were interpreted, discussed and summarized in a manuscript by L.M.S. and A.R.-I. with the support of all coauthors.

**Corresponding author**

Arash Rahimi-Iman: a.r-i@physik.uni-marburg.de

James C. Hone: jh2228@columbia.edu

**Authors' statement/Competing interests**

The authors declare no conflict of interest

**Additional information**

Supplementary Information accompanies this paper

**Figure Legends**

***Figure 1 | Schematics, micrograph of the sample and optical properties. a*** *Schematic exciton propagation in a 2D lattice with in-plane momentum* **q**$_{||}$ *and emission of photons with wave vector* ***k****, with the corresponding projection of detected angles in the Fourier space and corresponding in-plane dispersion relation.* ***b*** *Microscopy image of the h-BN-WSe$_2$-h-BN stack on an SiO$_2$/Si substrate. The sandwiched monolayer region is indicated by an arrow. Images taken during the stacking sequence and detailed markings are found in the Supporting Information. Inset: Schematic 2D heterostructure.* ***c*** *Optical spectrum of the encapsulated high-quality monolayer with exciton X$_0$, trion X\* and defect states D.* ***d*** *Exemplary k-space-resolved spectrum of a monolayer free-exciton mode at 15 K.*

***Figure 2 | Density-dependent dispersion changes of the WSe$_2$ exciton mode. a*** *and* ***b*** *depict energy–momentum-dispersion curves measured at 15 K in false-colour intensity representation in arbitrary units (linear colour scale from white to dark blue corresponding to minimum to maximum signal), for the excitation densities 0.1 and 3.6 kW/cm², respectively.* ***c*** *An overview on the density-dependent changes is given for the measurement range, showing a reduction of the curvature that corresponds to higher effective masses. Dashed lines represent fitted parabolic dispersion curves corresponding to the extracted effective masses labelling the data. Dispersion data is offset vertically by 1.1 meV for clarity. Error bars represent fit errors of Gaussian fits to the individual line spectra.*

***Figure 3 | Free-exciton emission characteristics in temperature-dependent optical dispersion measurements. a*** *Energy–momentum dispersion (false-colour intensity representation, similar to Fig. 2) for monolayer WSe$_2$ obtained at 15 K for quasi-resonant pumping. Similar data for 100 and 150 K is shown in* ***b*** *and* ***c****, respectively. For comparison to the corresponding quasi-resonant case in* ***b****, an emission spectrum measured at 100 K with off-resonant pumping is displayed in* ***d****.* ***e*** *summarizes the temperature series for quasi-resonant excitation at 15 K, 77 K, 100 K and 150 K. Error bars represent fit errors of Gaussian fits to the individual line spectra. Dashed lines represent fitted parabolic dispersion curves corresponding to the extracted effective masses labelling the data.*

## Supporting Information

# Shedding light on the monolayer-WSe$_2$ exciton's nature by optical effective-mass measurements

Lorenz Maximilian Schneider[1], Shanece S. Esdaille[2], Daniel A. Rhodes[2], Katayun Barmak[3], James C. Hone[2,*], and Arash Rahimi-Iman[1,*]

[1]Faculty of Physics and Materials Sciences Center, Philipps-Universität Marburg, Marburg, 35032, Germany

[2]Department of Mechanical Engineering, Columbia University, New York, NY 10027, USA

[3]Department of Applied Physics and Applied Math, Columbia University, New York, NY 10027, USA

### Stacking of the 2D heterostructure

The intermediary steps towards the completion of an h-BN-sandwiched WSe$_2$ monolayer used in the experiment are shown as optical micrographs in **Fig. SI.1**. Here, the identified h-BN buffer flake area is shown in **a**, on which the WSe$_2$ flake with monolayer section, as presented in **b**, has been isolated (**Fig. SI.1c** and **d**). Consecutively, an isolated h-BN flake for capping was identified (see **Fig. SI.1e**). The resulting h-BN-encapsulated monolayer stack placed onto the SiO$_2$/Si substrate is displayed in **Fig. SI.1f** and **g**. For better visibility, micrograph images in **Figs. SI.1d**, **e** and **g** are shown with high gamma value.

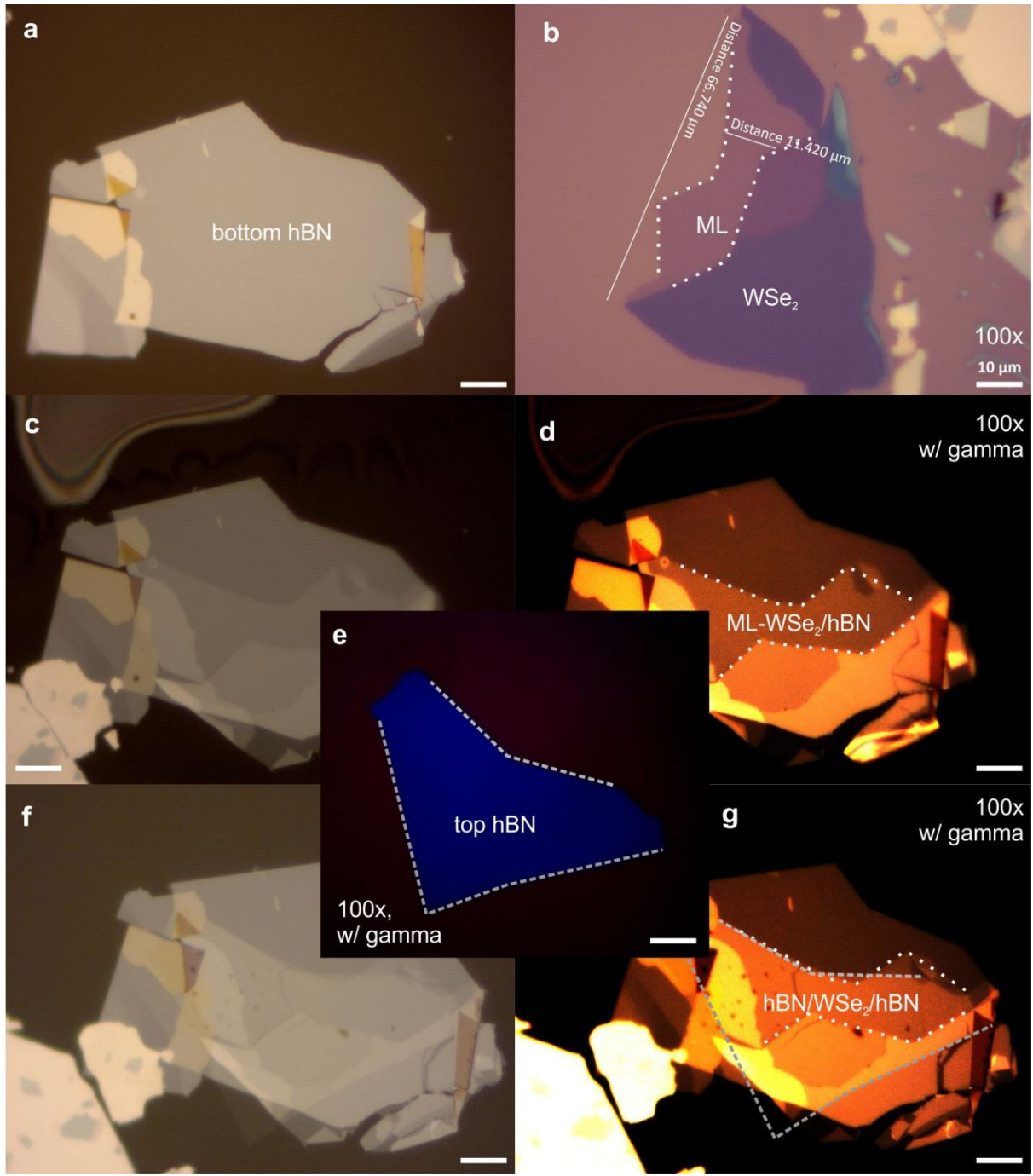

**Figure SI.1 | Sample preparation.** *Sequence of the stacking of a monolayer WSe$_2$ shown in **b** on a multi-layer thin h-BN flake seen in **a** before encapsulation and transfer to the substrate. The monolayer on h-BN is shown in **c** and **d** without and with gamma, respectively. **e** displays the top h-BN flake for encapsulation, as achieved in **f** and **g**. Dashed and dotted lines indicate the boundaries of the top h-BN and the sandwiched WSe$_2$ monolayer flake, respectively. Images were acquired using a 100x microscope objective.*

## Comparison pumping schemes

For quasi-resonant and off-resonant pumping, the results are very distinct, as **Figure SI.2** demonstrates. In the left plot, it can be seen that excitation with the Titanium-Sapphire laser at 693 nm (1.79 eV) results in a measurable dispersion at 15 K. In contrast, off-resonant excitation with a 445-nm diode laser at similar pump densities leads to a straight line in the Fourier-space-resolved emission spectrum at the same temperature level (10 K).

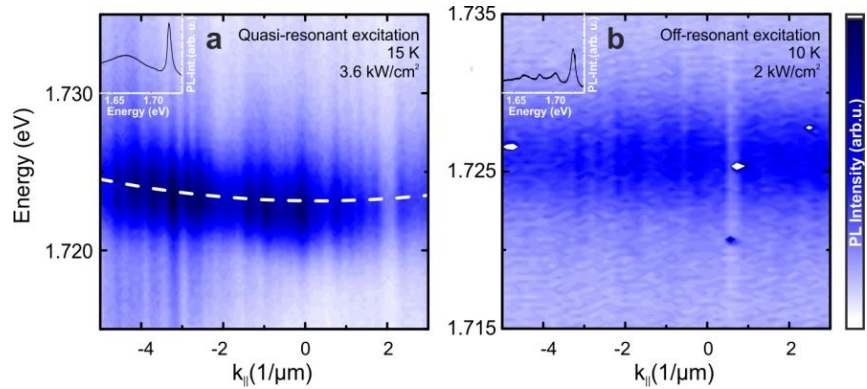

**Figure SI.2 | Pump-scheme-dependent Fourier-space spectra of the WSe$_2$ exciton mode.** *Dispersion curves measured at very low temperatures and for similar excitation densities are compared for quasi-resonant and off-resonant excitation (**a,** 15 K with 3.6 kW/cm² and, **b,** 10 K with 2 kW/cm², respectively). Dashed lines indicate the parabolic dispersion of a free exciton following the peak positions extracted for each line-spectrum individually by a Gauss fit. Insets represent corresponding integrated line spectra without angle-resolution.*